\documentclass[conference]{IEEEtran}
\IEEEoverridecommandlockouts
\usepackage{cite}
\usepackage{algorithmic}
\usepackage{textcomp}
\usepackage{xcolor}
\usepackage{graphicx}
\usepackage{multirow}
\usepackage{url}
\usepackage{comment}
\usepackage{todonotes}
\usepackage{amsmath, amsthm, amssymb, latexsym, amsfonts}
\usepackage{multicol}
\usepackage{subfig}
\usepackage{booktabs}
\usepackage{csquotes}
\usepackage{float}
\usepackage[hidelinks,colorlinks=true,linkcolor=blue,citecolor=blue]{hyperref}

\def\BibTeX{{\rm B\kern-.05em{\sc i\kern-.025em b}\kern-.08em
    T\kern-.1667em\lower.7ex\hbox{E}\kern-.125emX}}

\begin{document}

\title{Attention and Pooling based Sigmoid Colon Segmentation in 3D CT images}

\author{%
  Md Akizur Rahman\IEEEauthorrefmark{1},
  Sonit Singh\IEEEauthorrefmark{1},
  Kuruparan Shanmugalingam\IEEEauthorrefmark{1},
  Sankaran Iyer\IEEEauthorrefmark{1}, \\
  Alan Blair\IEEEauthorrefmark{1},
  Praveen Ravindran\IEEEauthorrefmark{2},
  Arcot Sowmya\IEEEauthorrefmark{1}
  \\
  \IEEEauthorblockA{%
    \IEEEauthorrefmark{1} School of CSE, University of New South Wales, Sydney, Australia
  }
  \IEEEauthorblockA{%
    \IEEEauthorrefmark{2} Department of Colorectal Surgery, Sydney Adventist Hospital, Sydney, Australia
  }
 
  Email: \url{md_akizur.rahman@unsw.edu.au} 
}

\maketitle

\begin{abstract}

Segmentation of the sigmoid colon is a crucial aspect of treating diverticulitis. It enables accurate identification and localisation of inflammation, which in turn helps healthcare professionals make informed decisions about the most appropriate treatment options. This research presents a novel deep learning architecture for segmenting the sigmoid colon from Computed Tomography (CT) images using a modified 3D U-Net architecture. Several variations of the 3D U-Net model with modified hyper-parameters were examined in this study. Pyramid pooling (PyP) and channel-spatial Squeeze and Excitation (csSE) were also used to improve the model performance. The networks were trained using manually annotated sigmoid colon. A five-fold cross-validation procedure was used on a test dataset to evaluate the network's performance. As indicated by the maximum Dice similarity coefficient (DSC) of $56.92 \pm 1.42$\%, the application of PyP and csSE techniques improves segmentation precision. We explored ensemble methods including averaging, weighted averaging, majority voting, and max ensemble. The results show that average and majority voting approaches with a threshold value of 0.5 and consistent weight distribution among the top three models produced comparable and optimal results with DSC of $88.11 \pm 3.52$\%. The results indicate that the application of a modified 3D U-Net architecture is effective for segmenting the sigmoid colon in Computed Tomography (CT) images.  In addition, the study highlights the potential benefits of integrating ensemble methods to improve segmentation precision.

\end{abstract}



\section{Introduction}~\label{Introduction}
\begin{figure}
    \centering
    \includegraphics[scale=0.37]{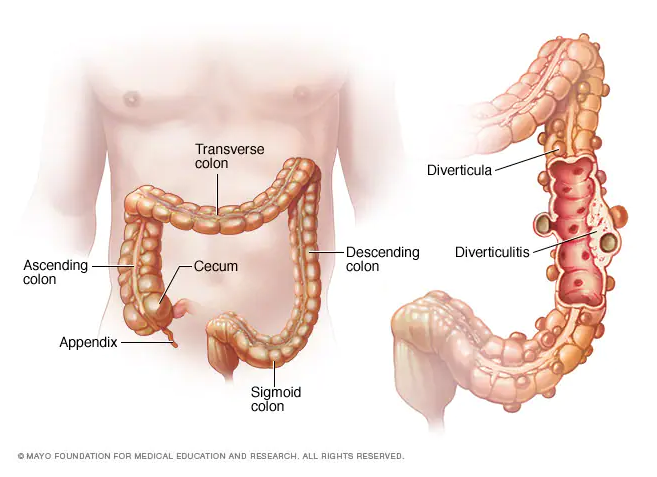}
    \caption{Colon Anatomy and diverticulitis. Figure taken from ~\cite{diverticulitis}}
    \label{fig:fig1}
\end{figure}

The \emph{sigmoid colon} is a segment of the large intestine that is situated in the lower left region of the abdomen. Abnormalities in the sigmoid colon, such as inflammation or tumours, can result in serious health complications, including diverticulitis and colon cancer. Timely identification and management of these ailments can potentially save lives. Diverticular diseases affect the colon and encompass a range of clinical conditions, from the mere presence of diverticula to complicated diverticulitis~\cite{Pecere:2020:acute_uncomplicated_diverticulitis}. Diverticula refer to small pouches that protrude from the inner lining of the digestive system. Diverticulitis occurs when the pouches become infected. Developing diverticulitis is associated with certain risk factors such as a diet that is high in fat and low in fibre, as well as low levels of physical activity. It is important to note that although these factors are linked with diverticulitis, not everyone who has these risk factors will necessarily develop the condition. The presence of air bubbles and gas outside the sigmoid colon, along with diverticulitis, during bleeding is often an indication of a more complex and severe medical condition. Based on these factors, it appears that the diverticulitis has progressed to advanced stages and may result in complications. Figure~\ref{fig:fig1} shows colon anatomy, having diverticulitis and its features outside on the sigmoid colon.

Medical imaging techniques, such as Computed Tomography (CT) scans, are frequently utilised to detect abnormalities in the colon~\cite{Guachi:2019:automatic_colorectal_segmentation,Gonzalez:2021:sigmoid_colon_segmentation}. Manually segmenting the sigmoid colon in medical images is a tedious and time-consuming process that demands specialised expertise. 

\begin{figure}
    \centering
    \includegraphics[scale=0.6]{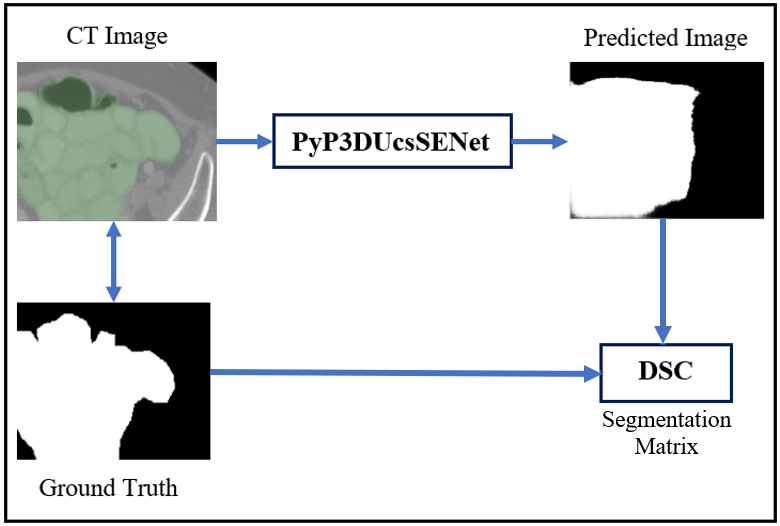}
    \caption{Block diagram showing sigmoid colon segmentation.}
    \label{fig:block_diagram}
\end{figure}

The precise segmentation of the sigmoid colon can aid in the detection of diverticula by determining their existence, position, and scope. The utilisation of this technique aids in the evaluation of inflammation, distinguishing it from other medical conditions, devising a treatment plan, and tracking the advancement of the disease. 

Automated segmentation of the sigmoid colon can be utilised as an alternative to the arduous and time-consuming manual task. Although there is lacking research on sigmoid colon segmentation~\cite{Gonzalez:2021:sigmoid_colon_segmentation}, recent studies have exhibited encouraging outcomes through the utilisation of Deep Learning methodologies, specifically Convolutional Neural Networks (CNNs) for medical image segmentation~\cite{Hu:2017:automatic_abdominal,Dou:2017:3D_deeply_supervised}. The field of computer-aided diagnosis is gaining popularity due to advancements in medical imaging and deep learning. The focus is on creating systems that can precisely segment various organs from medical images. According to recent studies~\cite{Zhou:2020:automatic_segmentation_of_multiple_organs,Dogan:2021:a_two_phase_approach}, automated segmentation systems possess the capability to improve diagnostic precision, decrease the workload on medical professionals, and ultimately enhance patient results. The development of deep learning-based sigmoid colon segmentation has encountered challenges despite some preliminary efforts~\cite{Gonzalez:2021:sigmoid_colon_segmentation}. Managing variations in shape, size, multi-curve structure, and location of the sigmoid colon between patients, as well as resolving noise and artefacts in medical images, are among these challenges. To address such barriers, developing sophisticated and robust deep-learning models capable of segmenting the sigmoid colon in medical images is necessary. This will result in improved patient diagnosis and treatment outcomes.
This study introduces an innovative deep learning architecture to segment the sigmoid colon. The architecture was built using 3D U-Net with Pyramid pooling (PyP) and channel-spatial Squeeze and Excitation (csSE), called PyP3DUcsSENet. An overview of the approach is in Figure~\ref{fig:block_diagram}. This research aims to evaluate and compare the segmentation performance of numerous 3D U-Net variants on 3D Computed Tomography (CT) images of the sigmoid colon. The proposed network was evaluated using a publicly available 3D CT images of colon cancer dataset. A comparative analysis was conducted to assess the performance of the proposed network in comparison to several variants of 3D U-Net. The results suggest that the PyP3DUcsSENet network is an efficient approach for segmenting the sigmoid colon. The PyP3DUcsSENet demonstrated superior precision and efficiency in comparison to alternative variants. Also, the ensemble technique results indicate that it is an effective approach for improving sigmoid colon segmentation performance. 

The rest of the paper is organised in the following Section II provides a comprehensive overview of the literature on medical image segmentation and colon segmentation. Then describes the employed materials and methods, including a thorough overview of the dataset, the proposed PyP3DUcsSENet architecture, and ensemble methods in Section III. Next, the experimental setup is detailed in Section IV. In Section V presents the study's findings and evaluates the performance of various variants of the 3D U-Net model utilising PyP3DUcsSENet and ensemble methodologies. Discusses the findings and their implications in Section VI, followed by recommendations for future research. Finally, in Section VII includes a summary of the paper's essential contributions and potential applications of this strategy.

\section{Related Work}
Medical image analysis requires accurate segmentation of various organs for diagnosis, treatment planning, and disease monitoring. Various techniques, such as patch-based approaches~\cite{Wang:2014:Geodesic_patch-based_segmentation}, multi-atlas techniques~\cite{Wolz:2013:automated_abdominal_multiorgan_segmentation}, and rule-based techniques~\cite{Chu:2013:multiorgan_segmentation,Park:2003:construction_of_an_abdominal}, have been proposed for organ segmentation. Although the methods have demonstrated satisfactory performance, segmenting extensive anatomical structures remains challenging. Deep learning techniques are gaining popularity for medical image segmentation as a result of their superior outputs and efficacy~\cite{Long:2015:Fully_convolutional_networks,Ronneberger:2015:U-Net,Estienne:2019:U-ResNet}. Using deep learning techniques, various organs including the liver~\cite{Hiemann:2009:comparison_and_evaluation,Dou:2017:3D_deeply_supervised,Li:2018:H-DenseUNet,Haq:2021:Liver_tumor_segmentation,Tran:2021:multi-layer_Unet,Singh:2023:radiological_diagnosis_of_CLD}, lung~\cite{Skourt:2018:lung_CT_image_segmentation,Asipong:2021:Coronavirus_infected_lung_CT}, pancreas~\cite{Huang:2021:semiautomated_DL_approach}, prostate~\cite{Hassanzadeh:2019:CNN_prostate_MRI,Shanmugalingam:2022:attention_guided}, and multiple organs~\cite{Hu:2017:automatic_abdominal,Wang:2019:abdominal_multi-organ_segmentatoin,Sultana:2020:automatic_multi-organ,Wu:2021:supervoxel_classification,Liu:2002:CT-based_multi-organ_segmenatation,Chen:2020:fully_automated_multiorgan_segmentation,Jia:2021:AMO-Net,Huang:2020:UNet3+,Zhou:2018:UNet++,Chu:2023:TSDNet} have been successfully segmented.

A comprehensive analysis of colon segmentation techniques was performed, encompassing approaches that do not rely on diverticulitis diseases. A method for colon segmentation was proposed in ~\cite{DeviK:2018:segmentation_of_colon}. The objective of the method is to identify polyps or growth and remove opacified fluid for virtual colonoscopy. The segmentation of air pockets was achieved through the utilisation of 3D Seeded Region Growing methodology. The seed was positioned in the axial slice that encompasses the rectum. The region growing process was executed using 6-connected neighbourhood connectivity. The segmentation of opacified fluid portions was achieved through the utilisation of fuzzy connectedness. The 3D rendering of the segmented colon was achieved by concatenating the segmented colon and opacified fluid. The study incorporated a total of 15 datasets, consisting of 7 datasets obtained from The Cancer Imaging Atlas (TCIA)~\cite{Cancerimagingarchieve} and 8 real-time datasets sourced from Imaging Centre, Coimbatore. The method proposed in this study attained a 98.73\% accuracy rate, indicating its efficacy in virtual colonoscopy for colon segmentation. In their study, Guachi et al. ~\cite{Guachi:2019:automatic_colorectal_segmentation} presented an automated technique for colon tissue segmentation. The technique is based on binary classification and utilises the LeNet-5 network, which was initially developed for handwritten digit recognition. In this study, the authors utilised the aforementioned technique to detect colon tissue at every individual pixel location. A CNN was trained using patches extracted from a dataset of CT images consisting of 100 slices. The convolutional LeNet-5 architecture proposal comprises convolution layers, ReLU activation functions, pooling layers, and fully connected layers. The method proposed in this study attained a colon tissue segmentation accuracy of $97.83\%$. This result indicates that the proposed approach is effective for colon tissue segmentation. In their study, Gonzalez et al. ~\cite{Gonzalez:2021:sigmoid_colon_segmentation} presented an iterative 2.5D deep learning methodology for the segmentation of the sigmoid colon. The approach employed a U-Net-like network structure. The approach employed a 3-D convolutional layer and down-sampling via max-pooling layers to process inputs slice by slice. Skip connections were utilised to facilitate the exchange of information between the descending and ascending branches of the network. The proposed methodology demonstrated encouraging results in both scenarios, i.e., with and without prior knowledge of other organs, for the automated segmentation of the sigmoid colon in CT images intended for radiotherapy treatment planning. The report is deficient in providing adequate information regarding their proposed architecture, thereby impeding the assessment of their approach's efficacy and replication of the outcomes.

The reviewed studies indicate that deep learning, specifically 3D U-Net variants, have the capability to perform precise and effective segmentation of medical images. Although 3D U-Net has demonstrated exceptional performance in medical image segmentation, there remain certain issues that necessitate further consideration. The complicated shapes of certain organs and their closeness to other organs continue to be challenging issues to solve. In addition, the presence of notable differences in the size and structure of organs may affect the precision of segmentation. Furthermore, there exists a requirement for enhancing the segmentation performance of tiny organs. It is recommended that future research endeavours focus on addressing the challenges associated with 3D U-Net in medical image segmentation and developing novel techniques to improve its overall performance.

\section{Materials and Methods}

We describe the dataset and different variants of the 3D U-Net model employed in this study. Specifically, we have made modifications to the network hyper-parameters and incorporated additional modules, including Pyramid pooling (PyP)~\cite{Shanmugalingam:2022:attention_guided} and channel-spatial Squeeze and Excitation (csSE)~\cite{Roy:2018:concurrent_spatial_channel} to enhance the performance of 3D U-Net  on sigmoid colon segmentation. Additionally, we  provide comprehensive information on the ensemble techniques employed in this study.

\subsection{Dataset}
\begin{figure}
    \centering
    \includegraphics[scale=1]{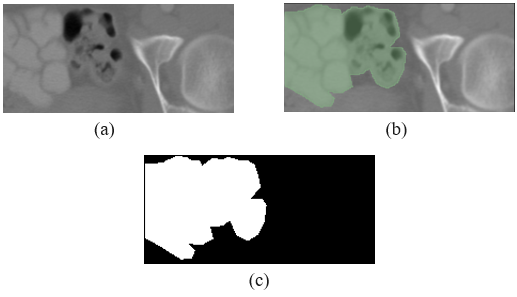}
    \caption{(a) Original abdominal CT slice. (b) Sigmoid colon annotation. (c) Ground-truth (annotated) mask.}
    \label{fig:sample_dataset}
\end{figure}

For this study, used a publicly available dataset of CT scans of colon cancer, collected from the medical segmentation decathlon challenge~\cite{medicaldecathlon}. At Memorial Sloan Kettering Cancer Centre in New York, $95$ CT scans were collected from patients having primary colon cancer resection. The scans were taken using an X-ray tube operating at $100-140$ kVp for $500-1782$ ms of exposure time at a current of $100-752$ mA. The reconstructed scans have diameters of between $274$ and $5000$ millimetres and slice thicknesses of between $1$ and $7.5$ millimetres. Although most of the cases in the dataset are those of colon cancer, we think it is a good place to begin our research with sigmoid colon segmentation in diverticulitis since the two diseases have many characteristics. Once CT scans were taken, the sigmoid colon was marked by hand on a 3D viewer. A colorectal expert from Liverpool Hospital checked the annotations for accuracy, and an example of this data is shown in Figure~\ref{fig:sample_dataset}.

\subsection{3D U-Net with Pyramid Pooling and Channel-Spatial Squeeze and Excitation (PyP3DUcsSENet)}
In the field of image segmentation, 3D U-Net, a CNN architecture, has seen extensive application~\cite{Guachi:2019:automatic_colorectal_segmentation,Gonzalez:2021:sigmoid_colon_segmentation}. It is an expansion of the 2D U-Net framework, optimised for dealing with 3D volumes. There are two nodes in the network: an encoder and a decoder. Multiple convolutional layers are used in the encoder, followed by a max pooling operation to decrease the input's spatial dimensions while simultaneously increasing the number of feature channels. To catch the complex feature details, this process is performed many times. In contrast, the decoder is built using up-convolutional layers that are concatenated with the relevant feature map from the encoder. The number of feature channels is decreased while the spatial dimensions of the input are increased by the up-convolutional layers. This allows the network to maintain the high-level characteristics learnt by the encoder while reconstructing the input image at its original resolution. As an efficient deep learning architecture for 3D medical image segmentation, the 3D U-Net model has shown state-of-the-art performance on several datasets. 

\begin{figure*}
    \centering
    \includegraphics[width=17cm, height=11cm]{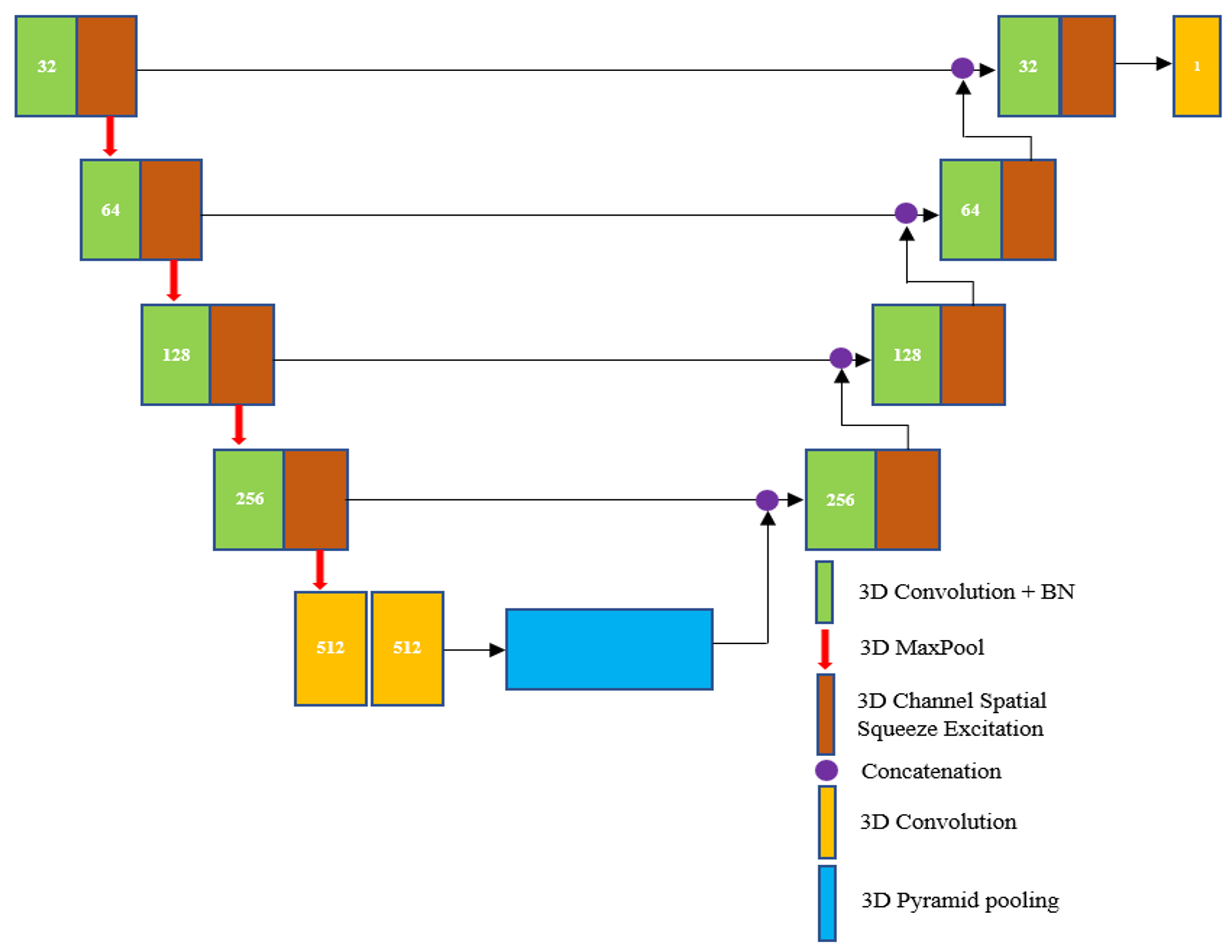}
    \caption{Architecture of the 3D U-Net with Pyramid Pooling and Channel-Spatial Squeeze and Excitation (PyP3DUcsSENet). The number of channels is denoted in the box.}
    \label{fig:architecture}
\end{figure*}

In this study, started with the fundamental 3D U-Net and investigated various modifications to its hyperparameters, including the number of filters in each layer, kernel size, dropout rate, and batch normalisation. In addition, we implemented two novel techniques known as Pyramid pooling (PyP) and channel-spatial Squeeze and Excitation (csSE) to boost performance. PyP is a technique that captures features with varying resolutions by aggregating at multiple dimensions. This procedure allows the network to extract finer details from the input images. csSE, on the other hand, is a technique that concentrates on improving the model's spatial and channel-wise attention. It enables the model to selectively emphasise the most pertinent features while suppressing the less significant ones, resulting in improved performance. This study achieved enhanced segmentation results for sigmoid colon data by implementing various improvements and strategies in the 3D U-Net model. The integration of csSE and PyP techniques significantly contributed to the improved performance. Figure~\ref{fig:architecture} shows the modified 3D U-Net architecture incorporating these enhancements.

\subsection{Ensemble methods}
This investigation employed ensemble techniques, wherein the three highest-performing 3D U-Net iterations were utilised to improve the precision of 3D CT image segmentation of the sigmoid colon. The objective was to integrate the prognostications of multiple models with the aim of enhancing the outcome beyond that of any individual model. This study conducted tests using the Average~\cite{Liu:2020:pancreas_segmentation}, Weighted Average~\cite{ShagaDevan:2022:weighted_average_ensemble}, Majority Voting~\cite{kavur:2020:basic_ensembles}, and Max Ensemble~\cite{kavur:2020:basic_ensembles} techniques, for a total of four ensemble techniques. The final segmentation was obtained by averaging the SoftMax values produced by each model on a given image, which is what the Average Ensemble does. To create the Weighted Average Ensemble, we first gave a value to each model's SoftMax output and then averaged the results. We utilised model weights of $0.33\%$, $0.33\%$, and $0.34\%$ for the top three models, respectively, based on their performance. We utilised the predicted binary masks from each model in the Majority Voting Ensemble to calculate the final segmentation by calculating the voxel with the greatest number of votes. To complete the segmentation, we used the Max Ensemble, which averaged the SoftMax output values for each voxel across all models. 


\section{Experimental Setup}

In this work, we evaluated various 3D U-Net architectures and we also introduce a unique architecture called PyP3DUcsSENet . In the latter, the analysis path includes consecutive $3 \times 3 \times 3$ convolutions followed by a ReLU activation function and a $2 \times 2 \times 2$ max pooling layer with strides in each dimension for each layer. The range of filters that are used is from $32$ to $512$. In the same way, in the synthesis path, each layer consists of a $2 \times 2 \times 2$ up-convolution, which is then followed by $3 \times 3 \times 3$ convolutions and a ReLU activation function. For the purpose of transferring high-resolution characteristics from the analysis path to the synthesis path, we utilised concatenation connections. In addition, in order to avoid the problem of overfitting, a dropout layer that had a probability of $0.4$ was included at the end of each ReLU that was included in each layer. In the last layer, we used a 1×1×1 convolution along with the sigmoid activation function to decrease the number of output channels to one. This number corresponds to the total number of labels. We utilised the Adam optimizer with a learning rate of $1e-4$ for all of the 3D U-Net variants and $1e-6$ for PyP3DUcsSENet. The momentum was set at $0.90$, and we ran the algorithm for $100$ iterations. During training, we used the Dice loss as our loss function, and DSC matrix was utilised to assess the final predicted segmentation outcomes.

The implementation of each of the 3D U-Net architectures was done in Python using the Keras framework, which is a high-level API for TensorFlow 2.1.0. The most powerful supercomputer in Australia, NCI Gadi, was used to carry out this studies. This machine allowed to access High Performance 24 CPUs and 2 GPUs (each Tesla V100 having 32 GB VRAM), together with 240 GB of Memory.

\section{Experimental Results}

A total of $80$ sigmoid colon annotated data samples were utilised, which were obtained from colon cancer datasets. Five-fold cross-validation was utilised for the segmentation procedure, using $64$ training samples and $16$ test samples in each fold. The experimental outcomes for several versions of the 3D U-Net are presented in Table~\ref{tab:table1}.

\begin{table}[]
    \centering
    \caption{Performance of six different architectures for sigmoid colon segmentation on 3D CT images along with the average DSC obtained from 5-fold cross validation.}
    \label{tab:table1}
    \begin{tabular}{l|c}
        \toprule
        \textbf{Model Name} & \textbf{Average DSC (\%) $\pm$ SD (\%)} \\
        \midrule
        3D U-Net & $45.60 \pm 3.13$ \\
        3D U-Net++ & $48.57 \pm 4.11$ \\
        3D wU-Net & $48.82 \pm 4.43$ \\
        3D ResU-Net & $55.33 \pm 1.79$ \\
        3D DenseU-Net & $55.61 \pm 1.70$ \\
        \textbf{PyP3DUcsSE-Net} & $\mathbf{56.92 \pm 1.42}$ \\
        \bottomrule
    \end{tabular} 
\end{table}

The outcomes of six different models that were trained for sigmoid colon segmentation in 3D CT images are shown in Table I, along with the average DSC values obtained from 5-fold cross-validation. With DSC scores of $55.33 \pm 1.79 (\%)$ and $55.61 \pm 1.70 (\%)$, respectively, the 3D Res-U-Net and 3D Dense U-Net models outperformed other fundamental 3D U-Net versions in terms of performance. A DSC of $56.92 \pm 1.42 (\%)$ was attained using the proposed PyP3DUcsSENet model, which significantly enhanced performance. Interestingly, all architectures did not perform any better after adding the PyP and csSE modules, except for the fundamental 3D U-Net model, which showed a modest increase in DSC.

 The results of the ensemble methods are recorded in Table~\ref{tab:table2}. Based on the DSC scores, we can see that the Average Ensemble and Majority Voting Ensemble techniques performed the best, with an average DSC score of $88.11 \pm 3.52(\%)$. 

\begin{table}[]
    \centering
    \caption{Performance of three models in ensemble for boosting sigmoid colon segmentation DSC on 3D CT images obtained from 5-fold cross validation.}
    \label{tab:table2}
    \begin{tabular}{l|c}
        \toprule
        \textbf{Ensemble Technique} & \textbf{Average DSC (\%) $\pm$ SD (\%)} \\
        \midrule
        \textbf{Average Ensemble} & $\mathbf{88.11 \pm 3.52}$ \\
        Weighted Average Ensemble & $88.00 \pm 3.61$ \\
        \textbf{Majority Voting Ensemble} & $\mathbf{88.11 \pm 3.52}$ \\
        Max Ensemble & $78.58 \pm 4.22$ \\
        \bottomrule
    \end{tabular} 
\end{table}

The Weighted Average Ensemble technique demonstrated favourable performance, exhibiting an average DSC score of $88.00 \pm 3.61(\%)$. In contrast, the Max Ensemble technique exhibited notably inferior performance, as evidenced by an average DSC score of $78.58 \pm 4.22(\%)$. This outcome suggests that the Max Ensemble technique may not be an appropriate approach for this particular task. This implies that the utilisation of the top three performing models results in an enhanced accuracy of sigmoid colon segmentation across all three techniques. In summary, the findings indicate that the utilisation of ensemble technique can serve as a valuable mechanism for enhancing the efficacy of segmentation models in the context of medical imaging applications. 


\section{Discussion}

\begin{figure*}
    \centering
    \includegraphics[scale=0.9]{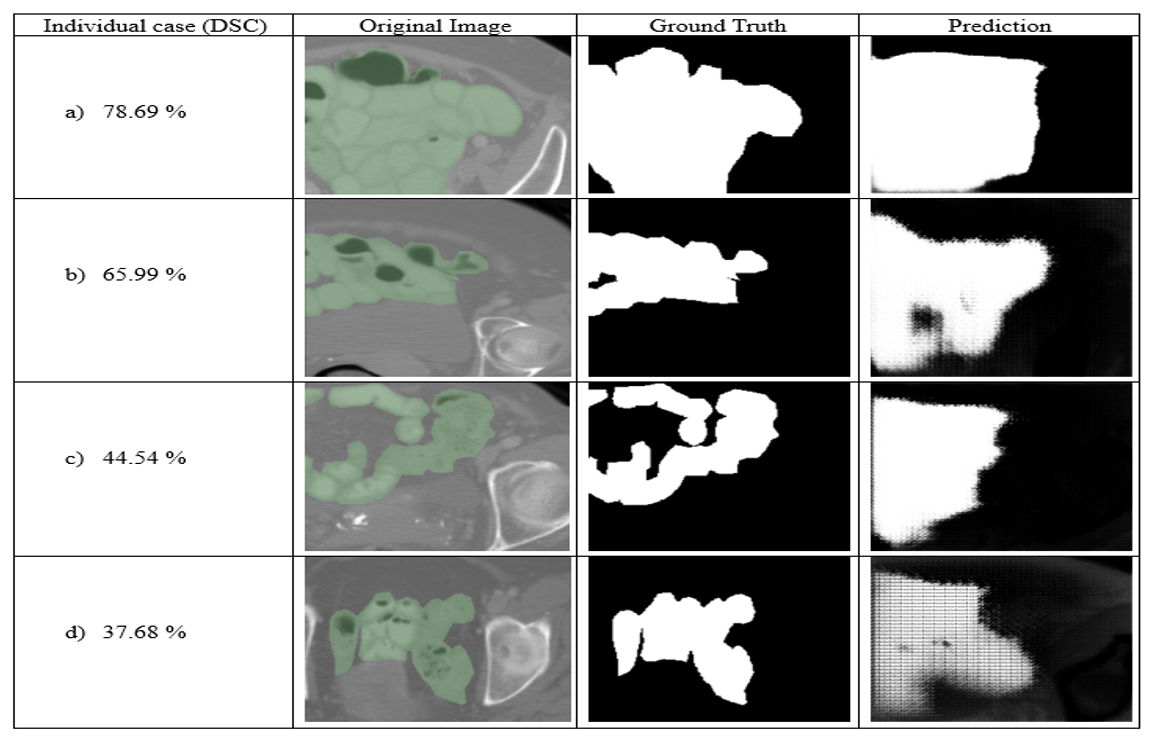}
    \caption{Individual cases visual comparison for the sigmoid colon segmentation.}
    \label{fig:figure5}
\end{figure*}

The results from PyP3DUcsSENet show varying levels of accuracy in sigmoid colon segmentation, as demonstrated in Figure~\ref{fig:figure5} for some of the cases. 

In Figure~\ref{fig:figure5}, case a), the predicted segmentation achieved a DSC score of $78.69\%$. This indicates a high level of accuracy because there was significant overlap observed between the predicted segmentation and the ground truth. In contrast, the predicted segmentation for case b) obtained a DSC score of $65.99\%$, indicating a moderate level of inconsistency with the ground truth. The DSC scores obtained for the predicted segmentations in cases c) and d) were notably lower, measuring $44.54\%$ and $37.68\%$, respectively. The scores show a notable decline in accuracy when compared to cases a) and b). This is because the predicted segmentations in cases c) and d) demonstrate significant differences in shape and size, resulting in noticeable discrepancies when compared to the actual ground truth. Figure 6 shows the average ensemble performance of both the best case and worst case.

\begin{figure*}
    \centering
    \includegraphics[scale=0.7]{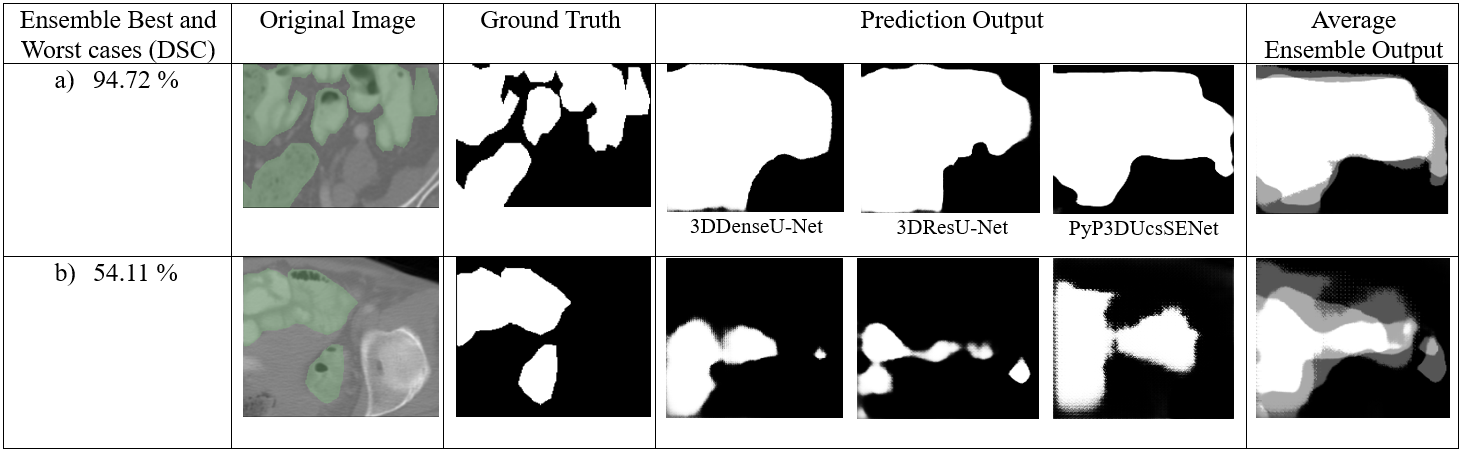}
    \caption{Ensemble performance of best and worst cases for sigmoid colon segmentation.}
    \label{fig:figure6}
\end{figure*}

In Figure~\ref{fig:figure6}, demonstrates a noteworthy enhancement in the performance of sigmoid colon segmentation due to the implementation of the ensemble model. In the best-case scenario (a), the ensemble model achieves an impressive DSC score of $94.72\%$, which indicates highly accurate segmentation. The ensemble model achieves a decent DSC score of $54.11\%$, even in the worst-case scenario. This observation implies that the network faces difficulties in learning the sigmoid colon efficiently because of the differences in its size and shape. The sigmoid colon is a relatively small and highly variable structure within the human body. The shape of the object exhibits a high degree of complexity and may exhibit considerable inter-patient variability. Insufficient manual annotation can lead to imprecisions and inconsistencies in the ground truth data, thereby exacerbating the difficulty of achieving accurate segmentation. The precision of segmentation of the sigmoid colon may exhibit significant variability, despite the utilisation of sophisticated deep-learning models. The results of this investigation demonstrate the existence of this diversity.

To improve the applicability of our segmentation model, forthcoming research will focus on expanding the dataset to include a broader spectrum of patient cohorts. Additionally, future studies will be directed towards improving the precision of manual annotation through the involvement of numerous specialists in the process of annotation. The implementation of these enhancements has the potential to yield segmentation outcomes that are more precise and reliable, rendering them particularly advantageous in medical settings.


\section{Conclusion}

The study presents the successful achievement of sigmoid colon segmentation in CT images through the utilisation of a modified 3D U-Net architecture. The modifications incorporated in the architecture encompassed the integration of pyramid pooling, channel-spatial squeezing and excitation algorithms, and pyramid pooling. The study demonstrates the potential benefits of employing ensemble methods to enhance precision, wherein the average ensemble and majority voting ensemble techniques yielded the most favourable results. The proposed method demonstrates the capacity to facilitate precise segmentation of the sigmoid colon.

\section*{Acknowledgement}
This research was facilitated by National Computational Infrastructure (NCI), which is supported by the Australian Government. We are also grateful for the financial support provided by The Australian Government Research Training Program (RTP) Scholarship at the University of New South Wales (UNSW) Sydney.

\section*{Declaration of competing interest}
The authors declare no conflicts of interest.

\bibliography{references.bib}
\bibliographystyle{IEEEtran}

\end{document}